\documentclass[twocolumn,amsmath,amssymb,floatfix,prl,showpacs,footinbib]{revtex4}
\usepackage{amsmath,amssymb,natbib,bm,graphicx,url,psfrag,times}

\usepackage{color}

\newcommand{\abs}[1]{\left|#1\right|}

\newcommand{\be}{\begin{equation}}
\newcommand{\ee}{\end{equation}}

\begin{document}
\title{Suppressing Charge Noise Decoherence in Superconducting Charge Qubits}
\author{J.\ A.\ Schreier}
\affiliation{Departments of Physics and Applied Physics, Yale University, New Haven, Connecticut 06520, USA}
\author{A.\ A.\ Houck}
\affiliation{Departments of Physics and Applied Physics, Yale University, New Haven, Connecticut 06520, USA}
\author{Jens Koch}
\affiliation{Departments of Physics and Applied Physics, Yale University, New Haven, Connecticut 06520, USA}
\author{D.\ I.\ Schuster}
\affiliation{Departments of Physics and Applied Physics, Yale University, New Haven, Connecticut 06520, USA}
\author{B.\ R.\ Johnson}
\affiliation{Departments of Physics and Applied Physics, Yale University, New Haven, Connecticut 06520, USA}
\author{J.\ M.\ Chow}
\affiliation{Departments of Physics and Applied Physics, Yale University, New Haven, Connecticut 06520, USA}
\author{J.\ M.\ Gambetta}
\affiliation{Departments of Physics and Applied Physics, Yale University, New Haven, Connecticut 06520, USA}
\author{J.\ Majer}
\affiliation{Departments of Physics and Applied Physics, Yale University, New Haven, Connecticut 06520, USA}
\author{L.\ Frunzio}
\affiliation{Departments of Physics and Applied Physics, Yale University, New Haven, Connecticut 06520, USA}
\author{M.\ H.\ Devoret}
\affiliation{Departments of Physics and Applied Physics, Yale University, New Haven, Connecticut 06520, USA}
\author{S.\ M.\ Girvin}
\affiliation{Departments of Physics and Applied Physics, Yale University, New Haven, Connecticut 06520, USA}
\author{R.\ J.\ Schoelkopf}
\affiliation{Departments of Physics and Applied Physics, Yale University, New Haven, Connecticut 06520, USA}

\begin{abstract}

We present an experimental realization of the transmon qubit, an improved superconducting  charge qubit derived from the Cooper pair box. We experimentally verify the predicted exponential suppression of sensitivity to $1/f$ charge noise [J.\ Koch et al., Phys.\ Rev.\ A \textbf{76}, 042319 (2007)].  This removes the leading source of dephasing in charge qubits, resulting in homogenously broadened transitions with relaxation and dephasing times in the microsecond range.  Our systematic characterization of the qubit spectrum, anharmonicity, and charge dispersion shows excellent agreement with theory, rendering the transmon a  promising qubit for future steps towards solid-state quantum information processing.
\end{abstract}
\date{December 20, 2007}
\maketitle

Over the last decade, superconducting qubits have gained substantial interest as an attractive option for quantum information processing, cf.\ Refs.\ \cite{makhlin,michel,you} for recent reviews.  Although there already exist different realizations of superconducting qubits \cite{nakamura,wallraff,wal,martinis}, all their coherence times are several orders of magnitude too short for large-scale quantum computation. 
Fortunately, an increase of coherence times from $2\,\text{ns}$ in the first superconducting qubit \cite{nakamura} to microsecond times in present experiments \cite{vion2,yoshihara,bertet3,wallraff2} has already been shown, giving rise to hope that the remaining gap can be overcome by optimized quantum circuits and better materials.
Coherence times can be either limited by dissipation ($T_1$) or dephasing ($T_2^*$). Most superconducting qubits have dephasing times much shorter than the limit $T_2^*=2T_1$ imposed by dissipation, because they are plagued by the influence of $1/f$ noise in charge, flux, or critical current. 
The transmon qubit is an improved design \cite{koch} derived from the original charge qubit \cite{bouchiat} that renders it immune to its primary source of noise, $1/f$ charge noise, without making it more susceptible to either flux or critical current noise.

The transmon consists of two superconducting islands connected by a Josephson tunnel junction. The tunneling of Cooper pairs  between the two islands is governed by two energy scales: the charging energy $E_C$ and the Josephson energy $E_J$. The transmon has a Hamiltonian identical to the Cooper pair box (CPB),
\be
\hat{H}=4E_C\left(\hat{n}-n_g\right)^2 - E_J \cos \hat{\varphi},
\label{CPB-gen}
\ee
where $\hat{n}$ denotes the number of excess Cooper pairs on one of the islands  and $n_g$ the offset charge due to the electrostatic environment. Because there are no dc connections to the qubit, $\hat{n}$ is integer-valued like an angular momentum, and the conjugate variable $\hat{\varphi}$ is a compact angle.
  Despite its basic CPB nature, the transmon is operated in a vastly different parameter regime where $E_J/E_C\gg 1$ (typically $E_J/E_C\sim50$). The primary benefit of this new regime is a suppression of the sensitivity to charge noise, which is exponential in the ratio $E_J/E_C$. The qubit spectrum becomes  more uniformly spaced in the transmon, 
but it has been shown in \cite{koch} that the anharmonicity in the spectrum only decays as a weak algebraic function of $E_J/E_C$, allowing it to be used as an effective two-level system. 
One of the reasons for the long coherence times of the design is that the state of the transmon qubit cannot be determined by any low-frequency measurement such as charge \cite{nakamura}, flux \cite{bertet3}, or quantum capacitance \cite{duty}.  Nevertheless, its large transition dipole makes it ideally suited for a dispersive circuit QED readout \cite{wallraff,blais1}.

Transmon qubits have already been employed successfully in several recent experiments \cite{schuster,houck1,majer}. In this paper, we present for the first time a detailed experimental characterization of the transmon itself, demonstrating the suppression of charge noise and resulting long coherence times. We have fabricated several transmon qubits, showing clean spectra in excellent agreement with the theoretical model, down to a few parts in $10^4$. 
We observe several energy levels of the transmon and show that the transition frequencies are distinct, with sufficient anharmonicity to allow us to perform fast control of the qubit.
Our measurements of the qubit frequency as a function of gate charge demonstrate the exponential suppression and hence immunity of the qubit to charge noise. Together, this enables coherent control of the qubit and results in a dephasing time $T_2^*$ exceeding $2\,\mu\text{s}$ without echo which approaches the limit $T_2^*=2T_1$.

We fabricate transmon qubits coupled to a transmission line cavity in a circuit QED architecture, realizing a Jaynes-Cummings Hamiltonian \cite{blais1,wallraff}. In this paper we present results from three qubits on two different samples, a low-$Q$ sample that enables fast spectroscopy and a high-$Q$ sample for the long coherence time measurements.
We employ minimal fabrication that limits materials complexity, requiring only two metal layers on a single-crystal sapphire substrate, with no crossovers or deposited dielectrics.  The cavity is made of $180\,\text{nm}$ thick dry-etched niobium, patterned with optical lithography, while the  qubits consist of double-angle evaporated aluminum ($100\,\text{nm}$ and $20\,\text{nm}$ thick layers), patterned with electron-beam lithography \cite{frunzio}. In our qubits, the superconducting islands are connected by a pair of junctions in parallel (each $0.18\,\mu\text{m} \times 0.25\,\mu\text{m}$), allowing the effective Josephson energy to be tuned by an external magnetic field, $E_J=E_J^\text{max}\abs{\cos(\pi\Phi/\Phi_0)}$.

\begin{figure}
    \centering
        \includegraphics[width=0.9\columnwidth]{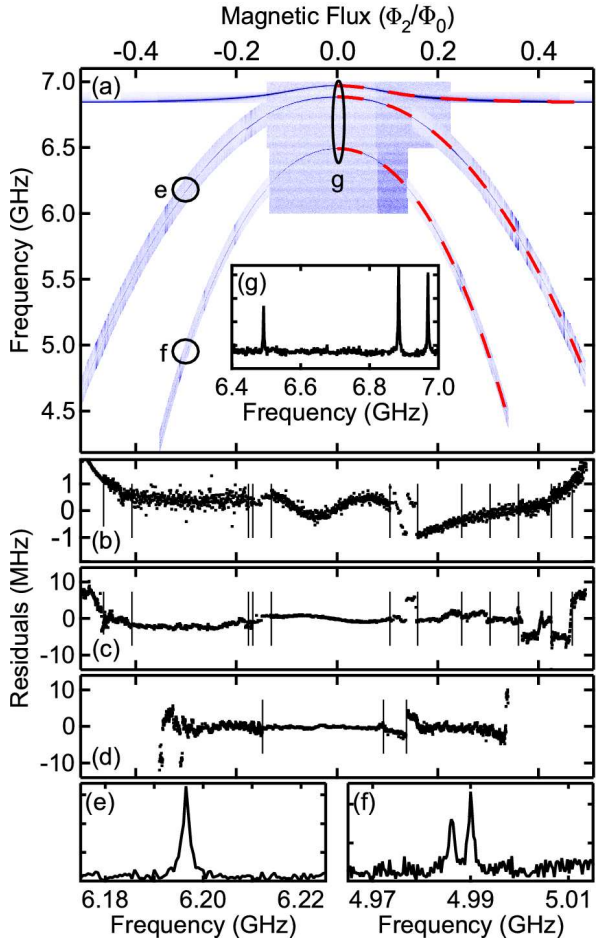}
    \caption{(Color online) Spectrum of two transmon qubits coupled to a transmission line cavity.
     (a) Observed spectrum of qubits and cavity as a function of magnetic field. Data obtained in several runs with both transmission and spectroscopy measurements, and manually stitched together [stitch points are given by vertical lines in (b)--(d)]. The dashed line is the theoretical fit to the data (for clarity only shown on the right). The charging energies $E_{C1}/h=386\,\text{MHz}$ and $E_{C2}/h=332\,\text{MHz}$ are obtained from the charge dispersion and held fixed. Fit parameters and their obtained optimal values are the maximum Josephson energies $E_{J1}^\text{max}/h=17.45\,\text{GHz}$, $E_{J2}^\text{max}/h=18.06\,\text{GHz}$, coupling strengths $g_1/2\pi=47\,\text{MHz}$, $g_2/2\pi=169\,\text{MHz}$, cavity frequency $\omega_c/2\pi=6.84\,\text{GHz}$, and the flux periodicity for each qubit with $\Phi_1=0.67\Phi_2$. (b)--(d) show the absolute errors of the model fit for the cavity, qubit 1, and qubit 2, respectively. (f) shows the only spurious avoided crossing, while (e) and (g) show typical line cuts from spectroscopy and transmission. 
\label{fig:fig1}}
\end{figure}

Varying the magnetic field allows for the measurement of the spectra for qubits and cavity over a wide frequency range. In Fig.\ \ref{fig:fig1}, we show data from the two-qubit sample, tracking the qubit and cavity frequencies as a function of the field. When the qubits are near resonance with the cavity, all three spectral lines can be observed in a transmission measurement. Away from resonance, the cavity is still measured in transmission, while the qubits are measured in spectroscopy \cite{wallraff,schuster2}. In the high-$E_J/E_C$ limit, the bare qubit frequency is expected to follow the asymptotic form $\hbar\omega_{01}=\sqrt{8E_JE_C} - E_C$ \cite{koch}. A more accurate description of the spectrum can be obtained from a full diagonalization of the transmon Hamiltonian \eqref{CPB-gen}, and using this as input to a two-qubit Jaynes-Cummings model.  The resulting theoretical fit agrees well with measurements of cavity and qubits over a frequency range of $2.5\,\text{GHz}$.

In order to allow for high fidelity qubit control, any qubit must not be coupled to uncontrolled degrees of freedom.
The complete spectrum enables a systematic search for spurious avoided crossings, often attributed to two-level fluctuators in junctions \cite{simmonds, martinis2}. 
We measure the spectra at $\sim2000$ magnetic field values such that each qubit never moves more than $2.5\,\text{MHz}$ in each field step. Data were taken in several independent runs, and small field offsets were needed to ``stitch"  separate runs together.
Each peak is fit to a Lorentzian to extract the center position.
The typical qubit line width is $2\,\text{MHz}$ due to power broadening \cite{schuster2}, with center frequency determined with $300\,\text{kHz}$ precision. Flux jumps, long-term flux drift and ``stitching" errors ultimately limit the overall accuracy to a few parts in $10^4$.

While magnetic flux  jumps are clearly visible in the spectrum, we see only one spurious crossing in one qubit (Fig.\ \ref{fig:fig1} at point c) with an avoided crossing of $3.8\,\text{MHz}$ at $\omega_{01}/2\pi=4.988\,\text{GHz}$, where each qubit is measured over a range of  $\sim2.5\,\text{GHz}$. This avoided crossing is observed to be local and flux-independent, meaning that it only affects one of the two qubits and is reproduced at both values of the magnetic flux where $\omega_{01}/2\pi=4.988\,\text{GHz}$.
With the given density and precision of the data set we can reliably detect all avoided crossing down to a splitting of $\sim4\,\text{MHz}$.  Requiring a hybridization of less than 1\% with the crossing, merely $38\,\text{MHz}$ of the presented two qubit spectrum is excluded, giving 99\% usability for qubit operations. Over most of the available frequency range, high-fidelity qubit control should be achievable, limited only by $T_1$ and $T_2^*$.

The density of avoided crossings we observe in this sample (one per five GHz per qubit) is typical of our devices.  Over the past several years, we have measured $\sim10$ transmon and $\sim10$ CPB qubits over a total frequency range of $\sim20\,\text{GHz}$, and seen three additional crossings with splittings of $8\,\text{MHz}$, $16\,\text{MHz}$ and $60\,\text{MHz}$.  With this low density of avoided crossings, the failure rate ($\lesssim 5\%$) of qubits due to the occurrence of a splitting at the desired operating frequency is acceptably small.  This may be due to the small area of the junctions, the simple fabrication process, and/or the well-controlled electromagnetic environment presented to the qubit by the cavity.

The ability to independently address the $01$ transition (where the integer numbers enumerate the transmon levels) is  a necessary requirement for reducing the transmon to an effective two-level system. Fig.\ 2 shows spectroscopic measurements of the 01 and 12 transitions, both of which are one-photon transitions and are well-resolved in frequency space. The 12 transition is measured while populating the first excited state with a separate drive on the 01 transition. The anharmonicity, $\alpha/2\pi=\nu_{12}-\nu_{01}=-455\,\text{MHz}$, is sufficiently large for fast (few ns) gate operations \cite{steffen2}. The $02$ transition can also be driven, but requires $35\,\text{dB}$ more power as it is a two-photon process.

\begin{figure}
    \centering
        \includegraphics[width=1.0\columnwidth]{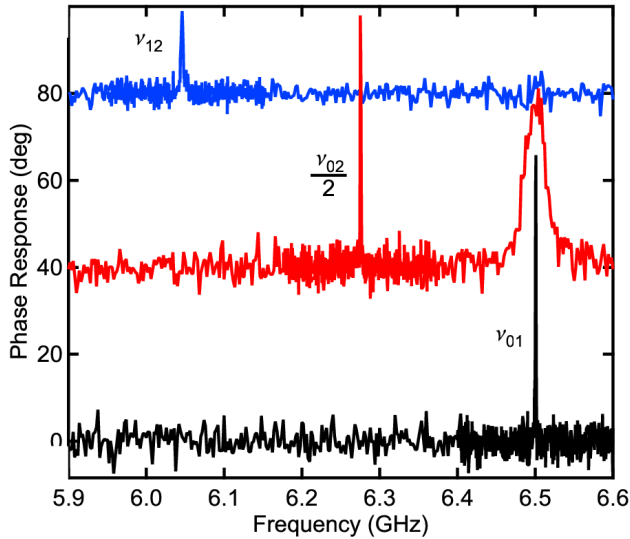}
    \caption{(Color online) Anharmonicity of a transmon qubit.  Data presented for qubit 1 from Fig.\ \ref{fig:fig1} at $E_J/E_C=40$.  Three qubit transitions are measured: the 01 transition in a single tone spectroscopic measurement (bottom curve), the two-photon 02 process with a $35\,\text{dB}$ stronger spectroscopy drive (middle curve, offset), and the 12 transition (top curve, offset) while populating transmon excited state with a second drive on the 01 transition.  The second excited state of the transmon is not populated with either the two-photon 02 process or a tone on the 12 transition at normal spectroscopy powers (bottom curve).  The 01 and 12 transitions are separated by $455\,\text{MHz}$; the transmon can therefore be treated as a two-level system even during fast control operations.   \label{fig:fig2}}
\end{figure}

Having established that the transmon can indeed be operated as a qubit, we now turn to the verification of the predicted immunity to $1/f$ charge noise. Charge noise causes random fluctuations in the qubit frequency and hence dephasing, limiting $T_2^*$ in the first CPB \cite{nakamura} to only a few nanoseconds. $T_2^*$ can be drastically improved by operating at an extremum of $\omega_{01}(n_g)$, referred to as a ``sweet spot" \cite{vion2} where the qubit is insensitive to first-order charge variations.
The crucial idea of the transmon is to drastically reduce the total variation in qubit frequency over all possible gate charges.
From the transmon Hamiltonian \eqref{CPB-gen}, one expects that this  ``charge dispersion" is suppressed as $\sim \exp(-\sqrt{8E_J/E_C})$ in the large-$E_J/E_C$ limit \cite{koch}.

\begin{figure}
    \centering
        \includegraphics[width=1.0\columnwidth]{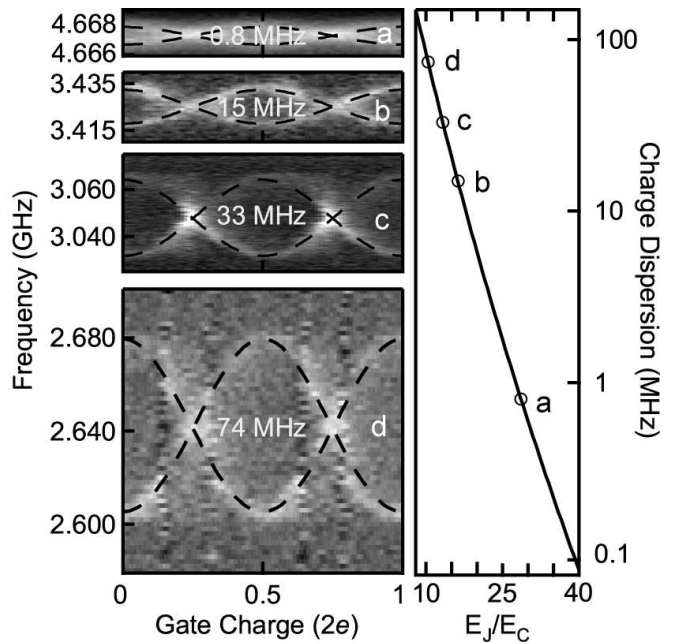}
    \caption{Exponential suppression of charge dispersion by tuning $E_J/E_C$.  Data presented for qubit 2 in Fig.\ \ref{fig:fig1} at four different values of $E_J$, resulting in (a) $E_J/E_C=28.6$, (b) $16.3$, (c) $13.3$, and (d) $10.4$.  Spectroscopic measurements of qubit frequency while changing a gate voltage reveal the expected sinusoidal frequency bands.  The total width of the band (charge dispersion) is decreased from $74\,\text{MHz}$ to $0.8\,\text{MHz}$ by changing $E_J/E_C$ from $10.4$ to $28.6$.  Two sinusoids are evident as random quasiparticle tunneling events cause the frequency curve to shift by one electron. The measured charge dispersion agrees well with the theoretical prediction, see right.\label{fig:fig3}}
\end{figure}

Here, we directly measure the suppression of charge dispersion for a transmon qubit in the crossover from the low-$E_J/E_C$ to the high-$E_J/E_C$ regime.
At several values of $E_J$, we determine the qubit frequency $\omega_{01}$ spectroscopically while varying the gate voltage for a single qubit with $E_{C2}=332\,\text{MHz}$.
As shown in Fig.\ \ref{fig:fig3}, this demonstrates the rapid decrease in charge dispersion when increasing $E_J/E_C$ and shows excellent agreement between experiment and theory.
As $E_J/E_C$ is increased by a factor of $2.5$, the charge dispersion is suppressed by two orders of magnitude.
At even higher values of $E_J/E_C$, we expect the charge dispersion for this qubit to be $13\,\text{kHz}$ at $E_J/E_C=50$ and $8\,\text{Hz}$ at $E_J/E_C=100$, essentially eliminating the effects of low-frequency charge noise.
The exponential sensitivity of charge dispersion also makes it a  useful tool for a very accurate determination of $E_C$. The charging energy obtained with this method is consistent with values extracted from the anharmonicity data.

From the Hamiltonian \eqref{CPB-gen}, we expect a single-valued function for $\omega_{01}$ which is sinusoidal in $n_g$. Instead, we observe a combination of two such curves shifted  by half a period, cf.\ Fig.\ \ref{fig:fig3}.
This is consistent with tunneling of quasiparticles between the two superconducting islands, resulting in the qubit frequency switching between the two curves. However, while a quasiparticle tunneling event completely dephases a qubit in the low-$E_J/E_C$ limit \cite{lutchyn1,lutchyn2} (``quasiparticle poisoning"), such events do not appreciably affect the frequency of the transmon and therefore have minimal contribution to pure dephasing for $E_J/E_C > 30$. These events still may cause a contribution to qubit relaxation; a rough estimate indicates that one in 20 tunneling events would cause the qubit to relax \cite{lutchyn1,lutchyn2,koch}. In the present experiment, individual tunneling events are too frequent to be resolved, i.e.\ the tunneling time $\tau$ is shorter than $50\,\text{ms}$. However, a separation of the two values of $\omega_{01}$ can be resolved spectroscopically down to $\sim1\,\text{MHz}$ [Fig.\ \ref{fig:fig3}(a)], implying no motional narrowing and $\tau \gtrsim 200\,\text{ns}$.

\begin{figure}
    \centering
        \includegraphics[width=0.9\columnwidth]{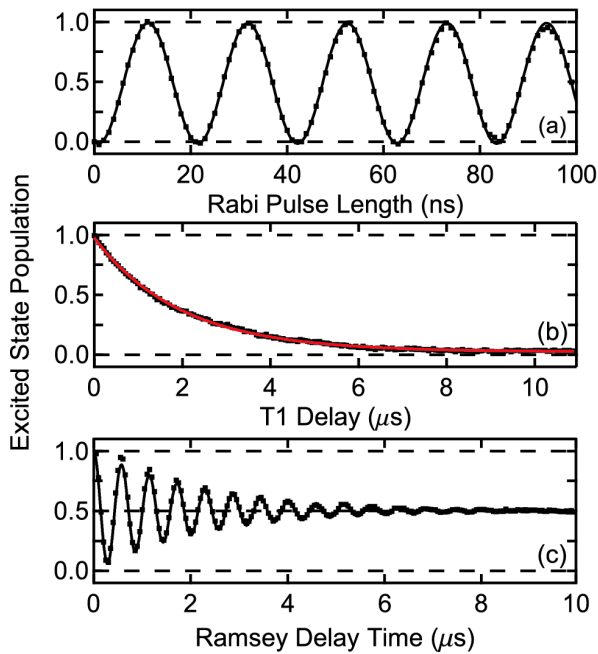}
    \caption{(Color online) High fidelity qubit control and long coherence times in a transmon qubit.  Data presented for a third qubit, measured at the flux sweet spot with $E_C\sim380\,\text{MHz}$ and $E_J/E_C \sim50$.  (a) Rabi oscillations with $100.5\%\pm 2\%$ visibility.  (b) Relaxation from qubit excited state.  Measurements of qubit state while varying delay after a $\pi$ pulse yield $T_1=1.87\,\mu\text{s}$. (c)  Ramsey fringes, measured without echo by varying a delay time between two $\pi/2$ pulses, demonstrate a long dephasing time $T_2^*=2.22\,\mu\text{s}$. \label{fig:fig4}}
\end{figure}

With this suppressed charge dispersion, the dephasing time for the transmon  is expected to be substantially improved. In Fig.\ \ref{fig:fig4}, we show time-domain measurements demonstrating reliable control and long coherence.  These experiments were performed with a third qubit in a higher-Q cavity to limit spontaneous emission due to the Purcell effect \cite{purcell,houck1}. The qubit was measured at the flux sweet spot \cite{vion2}, where $E_J/E_C\sim50$ and the residual charge dispersion is $\sim 15\,\text{kHz}$.

As a benchmark for the reliability of qubit control, we measure  the visibility of Rabi oscillations  to be $100.5\pm2$\%, calculated with linear extrapolation from a calibrating saturation pulse \cite{wallraff2}.  The relaxation time $T_1=1.87
\pm0.02\,\mu\text{s}$ is found by varying the delay after a $\pi$ pulse and fitting an exponential to the decay of the excited state population. The dephasing time $T_2^*$ is $2.22\pm0.03\,\mu\text{s}$, measured without echo by varying the Ramsey delay time between two $\pi/2$ pulses and fitting the observed fringes to an exponentially damped sinusoid. As $T_2^*$ is of the order of $T_1$, the qubit is nearly homogeneously broadened, even though 10 electrons of charge noise were intentionally applied to the gate.
The extracted pure dephasing time $T_\varphi=5.5\pm0.2\,\mu\text{s}$ is similar to the dephasing time expected from residual charge dispersion, $\sim10\,\mu\text{s}$.  Thus, we anticipate that future samples with even higher $E_J/E_C$ could further increase $T_\varphi$. Even away from the flux sweet spot, $T_2^*$ remains long without echo. In this sample we have measured $T_1=1.35\pm\,0.07\mu\text{s}$ and $T_2^*=1.75\pm0.05\,\mu\text{s}$ at $\Phi=0.23\Phi_0$ ($1\,\text{GHz}$) away from the flux sweet spot.

The dephasing times $T_2^*$ for both qubits on the two-qubit sample also approach a microsecond, though there $T_2^*$ is limited by $T_1$ due to the Purcell effect of the low-$Q$ cavity. 
In fact, short relaxation times in previous devices \cite{majer,houck1,schuster} can also be attributed to the Purcell effect, where relaxation due to higher harmonics of the cavity is essential for a proper prediction of $T_1$. 
A more detailed characterization of relaxation and dephasing demonstrating consistently long $T_1$ and $T_2^*$ across many qubits will be discussed in a future paper \cite{houck2}.

In conclusion, we have presented a detailed characterization of the transmon qubit, an optimized version of the CPB. The measurements were performed in a circuit QED architecture. All results show excellent agreement with theoretical predictions.  The qubits exhibit clean and well-understood spectra, with sufficient anharmonicity for fast qubit control.  The exponential suppression of charge noise sensitivity gives rise to nearly $T_1$ limited dephasing, giving hope that further improvements in $T_1$ could result in even longer $T_2^*$.

\begin{acknowledgments}
We acknowledge valuable discussions with John Martinis.  This work was supported in part by Yale University via a Quantum Information and Mesoscopic Physics Fellowship (AAH, JK), by CNR-Istituto di Cibernetica (LF), by LPS/NSA under ARO Contract No.\ W911NF-05-1-0365, and
the NSF under Grants Nos.\ DMR-0653377 and DMR-0603369. 
\end{acknowledgments}


\end{document}